\documentclass[aps,showpacs,superscriptadress,preprint]{revtex4}
\usepackage{graphicx}

\begin{document}
\title{Thermodynamics of system with density- and/or temperature-dependent mass particles}

\author{Shaoyu Yin$^1$ and Ru-Keng
Su$^{1,2}$\footnote{rksu@fudan.ac.cn}} \affiliation{
\small 1. Department of Physics, Fudan University,Shanghai 200433, China\\
\small 2. CCAST(World Laboratory), P.O.Box 8730, Beijing 100080, China\\
}

\begin{abstract}
The thermodynamics with medium effects expressed by the dispersion
relation of the temperature and density dependent particle mass is
studied. Many previous treatments have been reviewed. A new
thermodynamical treatment based on the equilibrium state is
suggested. Employing the quark mass density- and
temperature-dependent model, the discrepancies between our treatment
and others are addressed.
\end{abstract}


\maketitle

\section{Introduction}

It is generally accepted that the effective masses of particles will
change with temperature and density due to medium effects. Many
theoretical considerations, including the finite temperature QHD
model\cite{Walecka:1997}, Brown-Rho scaling\cite{Brown:1991}, QCD
sum rules\cite{Hatsuda:1993}, finite temperature QMC
model\cite{Song:1995,Saito:1997}, vacuum polarization Feynman
diagrams calculations\cite{Zhang:1997,Shiomi:1994,Wang:1999} and
\textit{etc.}, had been suggested to investigate the effective
masses of mesons and nucleons. By means of the Thermo-field
Dynamics, after summing the temperature- and density-dependent
vacuum polarization diagrams for $\pi$, $\sigma$, $\omega$, $\rho$
mesons and the three-line vertex diagrams which give the corrections
of the NN$\pi$, NN$\sigma$, NN$\omega$, NN$\rho$ couplings
respectively, we have extended the one-sigma\cite{Su:1990},
one-pion\cite{Qian:1993}, one-omega\cite{Su:1994} and one-rho
exchange potentials\cite{Zhang:1997,Wang:1999,Wang&Su:1999}, and
then the Bonn potentials\cite{Gao:1996} to finite temperature and
density. We have found that the attractive part of the NN
interaction becomes weaker and the repulsive part becomes stronger
when the temperature and/or density of the system increase. This is
of course very reasonable.

Besides theoretical study, many experimental results which predict
the changes of particle masses with temperature and density have
been shown. In particular, the experiments of TAGX collaboration
have shown directly that when the density of the nucleon medium
equals to $0.7n_0$ where $n_0$ is the saturation density, the
effective mass of nutral $\rho$-meson reduces to
610Mev\cite{Lolos:1998}. Both theoretical and experimental results
confirm that the medium effects are important for studying the
nuclear or quark systems.

To illustrate the medium effects more transparently in the theory
instead of the first principle calculation, many authors introduced
different hypothesis to represent the medium contributions to
particle masses, for example, introduced the density-dependent
vacuum energy $B(n)$ to modify QMC
model\cite{Jin:1996dual,Wang&Song:1999}, suggested the
density-dependent NN$\rho$ coupling to address liquid-gas phase
transition\cite{Qian:2000&2001}, supposed the $u$, $d$ and $s$ quark
masses depends on density to mimic the quark confinment
mechanism\cite{Fowler:1981}, introduced temperature-dependent vacuum
energy $B(T)$ to modify the quark mass density-dependent (QMDD)
model\cite{Zhang:2001}, and \textit{etc.}. Employing these
hypothesis, many physical properties of nuclear matter, quark
matter, nucleon system and hyperon system had been discussed.

Although the density- and temperature-dependent particle masses
$m^*(T,n)$ can mimic the medium effects, when we discuss the
thermodynamical behaviors of the system with such particles, many
difficulties will emerge. First, the dispersion relation for a
particle with energy $\epsilon$ and momentum $k$ becomes
\begin{equation}
\epsilon(k,T,n)=[k^2+m^*(T,n)]^{1/2}
\end{equation}
due to the medium effect. It was shown in
Ref.\cite{YangShinNan:1995} that the usual fundamental
thermodynamical partial derivative relation is not satisfied for
such system, we must add the additional terms. Secondly, it is well
known in thermodynamics that if we choose a pair of independent
variables, for example, temperature $T$ and volume $V$, and give the
suitable characteristic thermodynamical function, for example,
Helmholtz free energy $F=F(T,V)$, then all thermodynamical
quantities can be obtained by the partial derivatives of $F(T,V)$,
because
\begin{equation}
dF=-SdT-pdV,
\end{equation}
\begin{equation}
S=-(\frac{\partial F}{\partial T})_V,\qquad p=-(\frac{\partial
F}{\partial V})_T.
\end{equation}
Eq.(3) give the entropy and the equation of state respectively.
Other thermodynamical quantities such as internal energy $U$,
enthalpy $H$, Gibbs function $G$, heat capacities $c_V$ and $c_p$
can be obtained from Eq.(2) by derivatives. In above calculations,
no integral process will be needed and no integral constant has to
be determined. In fact, this is the reason why all calculation of
thermodynamical quantities in the canonical ensembles can be done
with the calculation of the partition function $Z=Z(T,V)$, because
the partition function $Z$ is related to free energy $F$ directly.
For systems with three independent variables, say, $T$, $V$ and
chemical potential $\mu$, the suitable characteristic function is
the thermodynamical potential $\Omega=\Omega(T,V,\mu)$ because
\begin{equation}
d\Omega=-SdT-pdV-Nd\mu.
\end{equation}
But for system with density-dependent mass particle, $\Omega$ is not
only a function of $T$, $V$, $\mu$, but also of the density $n$,
because $\Omega=\Omega(T,V,\mu,m^*(T,n))$. How to treat the
thermodynamics self-consistently is still a serious problem and has
many wrangles in present references. We will give a brief review of
various treatments and show their contradictory in the next section.
A few comments on different treatments will also be presented there.

This paper evolves from an attempt to suggest a new thermodynamical
treatment to study the system with density- and
temperature-dependent mass particles. Since the particle mass
depends on temperature and density, a lot of ambiguities will happen
when one uses the partial derivatives along a reversible process to
obtain the thermodynamical quantities. To avoid this difficulty,
instead of a \textit{reversible process}, we argue that we can
calculate the thermodynamical quantities at an \textit{equilibrium
state}. In fact, all the physical quantities, such as $p$, $U$, $F$,
$H$, $G$, are the functions of equilibrium state and have definite
values respectively. We can use the relations between
thermodynamical functions to obtain above quantities provided that
the thermodynamic potential $\Omega$ can be obtained. We will employ
the QMDD model to explain our treatment in detail in Sec.III. By
means of quark mass density- and temperature-dependent (QMDTD)
model, we will show our results and the comparison to other
treatments in Sec.IV. The last section is a summary.

\section{Different Thermodynamical treatments}
There have been several different kinds of treatments dealing with
the thermodynamics for the system with temperature- and/or
density-dependent mass particles. The basic difficulty comes from
Eq.(1) while one calculate the partial derivative of thermodynamical
functions along a reversible process. Though several authors had
still employed the usual direct formulae to calculate the
thermodynamical quantities and shown\cite{Chakrabarty:1989,
Chakrabarty:1991&1993}
\begin{equation}
P=-\frac{\Omega}{V},
\end{equation}
\begin{equation}
n_{i}=-\frac{1}{V}(\frac{\partial\Omega}{\partial\mu_{i}})_{T,n_{B}},
\end{equation}
\begin{equation}
\varepsilon=\frac{\Omega}{V}+\sum_{i}\mu_{i}n_{i}-\frac{T}{V}(\frac{\partial\Omega}{\partial
T})_{\{\mu_{i}\},n_{B}},
\end{equation}
where $n_B$ is the baryon density, $n_{i}$ is the number density of
particle $i$, $\mu_{i}$ is the corresponding chemical potential, and
$\varepsilon$ is the energy density. Their failure is obviously
because, with the temperature- and density-dependent mass term,
$\Omega$ is no longer an explicit function of $T$, $V$, $\mu_{i}$.
The usual direct partial derivative formulae are not applicable. To
overcome this difficulty, many different treatments or methods have
been suggested in the market, which can be generally categorized
into two kinds.

(I) The first kind of treatments takes the mass term as a function
of density and/or temperature and make a great effort to work
through the derivatives of a composite function along a reversible
process. In deriving the energy density and the pressure, extra
terms emerge in this approach due to the dependence of the mass on
the density and/or temperature. Although many authors work along
this direction to study the mass density and/or temperature
dependence, their formulae and results are very different with each
other.

(IA). For the the density-dependent mass, in
Ref.\cite{Benrenuto:1995dual}, they gave
\begin{equation}
P=-\frac{1}{V}(\frac{\partial(\Omega/n_{B})}{\partial(1/n_{B})})_{T,\{\mu_{i}\}}
=-\frac{\Omega}{V}+\frac{n_{B}}{V}(\frac{\partial\Omega}{\partial
n_{B}})_{T,\{\mu_{i}\}},
\end{equation}
\begin{equation}
\varepsilon=\frac{\Omega}{V}-\frac{n_{B}}{V}(\frac{\partial\Omega}{\partial
n_{B}})_{T,\{\mu_{i}\}}+\sum_{i}\mu_{i}n_{i}-\frac{T}{V}(\frac{\partial\Omega}{\partial
T})_{\{\mu_{i}\},n_{B}},
\end{equation}
where $n_i$ still satisfies Eq.(6). The extra terms in Eqs.(8, 9)
produce significant changes in the energy density $\varepsilon$, and
makes that the pressure become negative in low density regions.

(IB). In Ref.\cite{Peng:1999,Peng:2000}, for the density-dependent
mass, the pressure and the energy density become
\begin{equation}
P=-\frac{\Omega}{V}+\frac{n_{B}}{V}(\frac{\partial\Omega}{\partial
n_{B}})_{T,\{\mu_{i}\}},
\end{equation}
\begin{equation}
\varepsilon=\frac{\Omega}{V}+\sum_{i}\mu_{i}n_{i}-\frac{T}{V}(\frac{\partial\Omega}{\partial
T})_{\{\mu_{i}\},n_{B}}.
\end{equation}
They use the same pressure formula as that of (IA) Eq.(8), but do
not agree with the expression in Eq.(9), because it cannot give a
correct QCD vacuum energy. Then they use the same energy density
formula as Eq.(7).

(IC). In Ref.\cite{Wen:2005}, for the density- and
temperature-dependent mass, the pressure and the energy density read
\begin{equation}
P=-\widetilde{\Omega}-V\frac{\partial\widetilde{\Omega}}{\partial
V}+n_{B}\sum_{i} \frac{\partial\widetilde{\Omega}}{\partial
m_{i}}\frac{\partial m_{i}}{\partial n_{B}},
\end{equation}
\begin{equation}
\varepsilon=\widetilde{\Omega}-\sum_{i}\mu_{i}
\frac{\partial\widetilde{\Omega}}{\partial\mu_{i}}-
T\frac{\partial\widetilde{\Omega}}{\partial
T}-T\sum_{i}\frac{\partial\widetilde{\Omega}} {\partial
m_{i}}\frac{\partial m_{i}}{\partial T},
\end{equation}
where $\widetilde{\Omega}=\Omega/V$ stands for the density of the
thermodynamical potential.

(II) The second kind of treatments add different terms to the system
to keep the usual thermodynamical derivative relations unchanged.
For example, in order to make the Eq.(4) and the calculation of the
thermodynamical potential self-consistently, they add extra terms to
the Hamiltonian or to the thermodynamical potential. These
additional terms differ for different authors.

(IIA). To keep the fundamental thermodynamic relations
\begin{equation}
\varepsilon(T)=T\frac{dp(T)}{dT}-p(T),\qquad s=(\frac{\partial
p}{\partial T})_{\mu},\qquad n=(\frac{\partial p}{\partial \mu})_{T}
\end{equation}
self-consistently, in Ref.\cite{YangShinNan:1995}, a term
$E^{*}_{0}$ had been added to the Hamiltonian. This term is
determined by the condition
\begin{equation}
(\frac{\partial p}{\partial c_{i}})_{T,\mu,\{c_{j\neq i}\}}=0,
\end{equation}
where $\{c_{i}\}$ is the temperature- and density-dependent terms in
the Hamiltonian. Then the pressure and energy density become
\begin{equation}
p(T,\mu,\{c_{i}\})=\mp\frac{gkT}{2\pi^2}\int k^2dk\ln[1\mp
\exp^{-\beta(\epsilon(k)-\mu)}]-B^*,
\end{equation}
\begin{equation}
\varepsilon(T,\mu,\{c_{i}\})=\frac{g}{2\pi^2}\int\frac{\epsilon(k)k^2
dk}{\exp^{-\beta(\epsilon(k)-\mu)}\mp1}-B^*,
\end{equation}
where
\begin{equation}
B^*=\lim_{V\rightarrow\infty}\frac{E_0^*}{V},
\end{equation}
and $g$ is the degeneracy factor.

(IIB). In Ref.\cite{Wang:2000}, an extra term $\Omega_{a}(n_{B})$
had been added to the thermodynamical potential,
\begin{equation}
\widetilde{\Omega}=-\sum_{i}\frac{g_{i}T}{(2\pi)^3}\int dk^{3}
\ln(1+e^{-\beta(\epsilon_{i}(k))-\mu_{i}})+\Omega_{\alpha}(n_{B}),
\end{equation}
where $\Omega_{a}(n_{B})$ is determined by the constraint
\begin{equation}
\frac{\partial\widetilde{\Omega}}{\partial n_B}|_{\{\mu_{i}\}}=0.
\end{equation}
At zero temperature, the corresponding thermodynamical formulae
become
\begin{equation}
p=-\widetilde{\Omega},
\end{equation}
\begin{eqnarray}
\varepsilon&=&\sum_{i}\frac{g_{i}}{48\pi^{2}}\{\mu_{i}[\mu_i^2-m_i^2(n_B)]^{1/2}
[6\mu_i^2-3m_i^2(n_B)]\nonumber\\&&-3m_i^4(n_B)\ln[\frac{\mu_i+
[\mu_i^2-m_i^2(n_B)]^{1/2}}{m_i(n_B)}]\}+\Omega_{\alpha}(n_{B}),
\end{eqnarray}
\begin{equation}
n_i=\frac{g_i}{6\pi^2}[\mu_i^2-m_i^2(n_B)]^{3/2}.
\end{equation}

In summary, from above treatments we come to a conclusion that how
to treat the thermodynamics with the medium effect is still a
serious problem and has made many wrangles in present
references\cite{Chakrabarty:1989,Chakrabarty:1991&1993,Benrenuto:1995dual,Peng:1999,Peng:2000,Wen:2005}.
In fact, many treatments are contradict with each other. To show
their confusion and ambiguity, we give a few comments in the
following:

(1). The first argument is that the dispersion relation Eq.(1) and
the derivative formula of thermodynamical potential Eq.(4)
correspond to different conditions. In Eq.(1), the medium effect has
been taken into account. This effect makes that the particle mass
becomes a temperature- and density-dependent effective mass. But
Eq.(4) corresponds to the thermodynamical system only where
$\Omega$, $S$, $T$, $p$ are the thermodynamical quantities of the
system respectively. They do not include the medium effect. In order
to make Eq.(1) and the derivative formula of thermodynamical process
self-consistent, instead of Eq.(4), we must establish a formula of
thermodynamical process for the total system. It includes the
variations of the quantities not only for the system, but also for
the medium. In this formula, all extensive quantities must be
replaced by the quantity of the total system, for example,
$S\rightarrow S_s+S_m$, where $S_s$ and $S_m$ are the entropy of the
system and the medium respectively. The problem is that so far we do
not know how to calculate $S_m$ and other extensive thermodynamical
quantities of the medium directly.

(2). To illustrate above argument more transparently, we imagine two
systems. These two systems are almost identical: their only
difference is that one system consists of constant mass ($m_1$)
particles while the other of temperature- and density-dependent mass
($m_2(T,n)$) particles. Fixed the temperature and density as $T_0$
and $n_0$ respectively, which satisfy
\begin{equation}
m_2(T_0,n_0)=m_1,
\end{equation}
at the equilibrium state with $(T_0,n_0)$, obviously, the two system
are completely the same. They arrive at and stay in the same
equilibrium state and have the same thermodynamical quantities, such
as the pressure and the energy density, which are just functions of
equilibrium state. But if we use equations with partial derivatives
to calculate these quantities of the later system, extra terms, such
as $\frac{\partial \widetilde{\Omega}}{\partial m_i} \frac{\partial
m_i}{\partial n_B}$ and $\frac{\partial \widetilde{\Omega}}{\partial
m_i} \frac{\partial m_i}{\partial T}$ in Eqs.(12, 13), emerge. These
terms will never appear in the calculation of the former system
since in this system $m_1$ is a constant. This simple example tells
us that the first kind of treatments is not right because the medium
effect has been neglected in Eq.(4).

(3). Now we hope to give a brief comment on the second kind of
treatments. They hope to add an extra term to the system to consider
the medium effect. But unfortunately, in general, the extra term
cannot be determined by an additional constraint in terms of partial
derivatives uniquely. It cannot be expressed as a zero value of
thermodynamical functions, because the effective mass $m^*(T,n)$
comes from the interaction of the particle and the medium. This
interaction depends on temperature and density. As an example, let's
discuss the treatment (IIB). Obviously, if we add an arbitrary
temperature function $f(T)$ to $\Omega_\alpha(n_B)$, the new
expression for total $\widetilde{\Omega}$ still satisfies the
constraint Eq.(20), but the additional arbitrary function of $T$
will change the thermodynamical functions, such as the entropy,
which depend on the derivation of temperature.

\section{New Treatment Based on the Equilibrium State}
The above arguments impress us to give up the derivative calculation
of thermodynamical quantities along a reversible process, since
Eq.(4) dose not include the medium effect. Instead of studying the
reversible process, we focus our attention on equilibrium states. We
suggest a new treatment which is based on the equilibrium state in
this section. We will show there is no ambiguity in our treatment.

According to the thermodynamics of equilibrium state, all
thermodynamical functions such as $\Omega$, $S$, $U$, $G$, $p$ ...
have definite values respectively at an equilibrium state. If we can
calculate the thermodynamical potential $\Omega$ and the Gibbs'
function $G$ for a fixed equilibrium state, we can use the following
definitions to find other thermodynamical quantities:
\begin{eqnarray}
U&=&\sum_in_i\epsilon_i,\\
n&=&\frac{N}{V}=\sum_in_i=\frac{g_i}{e^{\beta(\epsilon_i-\mu)}\pm
1},\qquad (+\ for\ Fermion\ and\ -\ for\ Boson)\\
G&=&\sum_in_i\mu_i\\
F&=&\Omega+G,\\S&=&\frac{U-F}{T},\\P&=&-\frac{\Omega}{V}.
\end{eqnarray}

To show our treatment explicitly, we employ the QMDTD model
\cite{Zhang:2001,Zhang:2002,Zhang:2003,Wu:2005,Mao:2006}. The QMDTD
model is extended from the QMDD model which was first suggested by
Fowler \textit{et al.} many years ago \cite{Fowler:1981}. According
to the QMDD model, the masses of $u$, $d$ quarks and $s$ quark are
given by
\begin{equation}
m_{q}=\frac{B}{3n_B},\qquad(q=u,\overline{u},d,\overline{d}),
\end{equation}
\begin{equation}
m_{s,\overline{s}}=m_{s0}+\frac{B}{3n_B},
\end{equation}
where $B$ is the vacuum energy density and $m_{s0}$ is the current
mass of the strange quark. It is clear that the masses of quarks
become infinity when the baryon density goes to zero, which means
that the quark confinement in this model is permanent. To remove the
permanent confinement of quark, we modified the QMDTD model by
introducing \cite{Zhang:2001,Zhang:2002}
\begin{equation}
m_{q}=\frac{B(T)}{3n_B},\qquad(q=u,\overline{u},d,\overline{d}),
\end{equation}\begin{equation}
m_{s,\overline{s}}=m_{s0}+\frac{B(T)}{3n_B},
\end{equation}
where
\begin{equation}
B(T)=B_{0}[1-(\frac{T}{T_{c}})^2].
\end{equation}
The quark mass depends on both density and temperature, so it is
called a QMDTD model. We have used this model to study the
properties of strange quark matter (SQM)
\cite{Zhang:2002,Zhang:2003,Wu:2005,Mao:2006}.

For the system of SQM, in which the masses of quarks satisfy
Eqs.(33-35), at equilibrium state, the thermodynamical potential of
the system reads \cite{Benrenuto:1995dual,Wen:2005,Zhang:2002}
\begin{equation}
\Omega=-\sum_{i}\frac{g_{i}TV}{(2\pi)^3}\int d^3k\
\ln(e^{-\beta(\epsilon_{i}(k)-\mu_{i})}+1),
\end{equation}
The internal energy $U$ and the Gibbs' function $G$ are
\begin{equation}
U=\sum_{i}g_{i}\int d^3k
\frac{\epsilon_{i}(k)}{e^{-\beta(\epsilon_{i}(k)-\mu_{i})}+1},
\end{equation}
\begin{equation}
G=\sum_{i}g_{i}\int d^3k\frac{\mu_{i}}{
e^{-\beta(\epsilon_{i}(k)-\mu_{i})}+1},
\end{equation}
respectively. Other thermodynamical quantities can be obtained by
Eqs.(25-30).

At finite temperature, the antiquarks must be considered. The baryon
density satisfies
\begin{equation}
n_{B}=\frac{1}{3}(\Delta n_{u}+\Delta n_{d}+\Delta n_{s}),
\end{equation}
where
\begin{equation}
\Delta n_{i}=n_{i}-n_{\overline{i}}=\frac{g_{i}}{(2\pi)^{3}}\int
d^{3}k (\frac{1}{\exp[\beta(\epsilon_{i}(k)-\mu_{i})]+1} -
\frac{1}{\exp[\beta(\epsilon_{i}(k)+\mu_{i})]+1}),
\end{equation}
($n_{\overline{i}}$)$n_i$ is the number density of the (anti)flavor
$i (i=u,s,d)$, $g_i=6$, for antiquark $\mu_{\overline{i}}=-\mu_i$,
Following Ref. \cite{Chakrabarty:1991&1993,Zhang:2002}, the system
of SQM must satisfy the constraints
\begin{equation}
\mu_s=\mu_d=\mu_u+\mu_e,
\end{equation}
because inside SQM, $s$ (and $\overline{s}$) quarks are produced
through the weak process
\begin{equation}
u+d\leftrightarrow u+s,\qquad s\rightarrow
u+e^{-}+\overline{\nu}_{e},\qquad d\rightarrow
u+e^{-}+\overline{\nu}_{e},\qquad u+e^{-}\rightarrow d+\nu_{e},
\end{equation}
and similarly for antiquarks. The condition of charge neutrality
reads
\begin{equation}
2\Delta n_{u}=\Delta n_{d}+\Delta n_{s}+3\Delta n_{e}.
\end{equation}

\section{Results and Discussion}
For the convenience of comparison, our numerical calculations have
been done by adopting the parameters $B_0=170$MeVfm$^{-3}$,
$m_{s0}=150$MeV and $T_c=170$MeV, as that of Ref.\cite{Zhang:2002}.
Our results are summarized in Figs.1-4 and Table I.

The energy per baryon vs. baryon number density $n_B$ is shown in
Fig.1 where the temperature is fixed at $50$MeV. In Fig.1, the solid
line refers to the present treatment and other four dashed lines
refer to treatments in
Ref.\cite{Chakrabarty:1989,Chakrabarty:1991&1993}, IA, IB and IC
respectively. We see the solid line is lower than the others. The
saturation points for different treatments are summarized in Table
I. We find from Fig.1 and Table I that the differences are
remarkable for different treatments.
\begin{table}[b]
\begin{tabular}[t]{c|cc}
\hline\hline
\centering
treatments & $n_{B0}(fm^{-3})\qquad$ & $\varepsilon/n_B$ \\
\hline
in refs.[21,22]& $0.46\qquad$ & $1023$\\
IA & $0.55\qquad$ & $1083$\\
IB & $0.36\qquad$ & $1120$\\
IC & $0.34\qquad$ & $1084$\\
our new treatment & $0.45\qquad$ & $1007$\\
\hline
\end{tabular}\label{table1}
\caption{Saturation points at $T=50$MeV for different treatments.}
\end{table}

To illustrate the thermodynamical characters of our treatment, we
show the internal energy and the free energy curves calculated by
our treatment in Fig.2 respectively. The energy per baryon
$\varepsilon/n_B$ (solid lines) and the free energy per baryon
$f/n_B$ (dashed lines) vs. baryon density for different temperatures
$T=0, 30, 50, 100$ MeV are shown in Fig.2. We find that
$\varepsilon/n_B$ increases and $f/n_B$ decreases as the temperature
increases. The saturation points exist for all temperatures.

The equation of state for different treatments are shown in Fig.3,
where the solid line refers to present treatment and other lines for
previous treatments, respectively, as labeled in the figure. The
solid line exhibits a significant property which differs from that
of the lines of treatments IA, IB and IC. The pressure is definitely
positive in our treatment. But for IA, IB and IC treatments, the
pressure becomes negative in the small energy density regions. The
pressure will not be negative due to its thermodynamical treatment
for a system with positive energy. This result confirms that our
treatment is correct.

In Ref.\cite{Wen:2005}, the authors claimed that the dispersion
relation of effective mass $m^*$ must satisfy the constraint
\begin{equation}
\lim_{T\rightarrow 0}\frac{\partial m^*}{\partial T}=0,
\end{equation}
based on their formula for entropy
\begin{equation}
s=-\frac{\partial \Omega}{\partial
T}-\sum_i\frac{\partial\Omega}{\partial m_i}\frac{\partial
m_i}{\partial T}.
\end{equation}
If $\lim_{T\rightarrow 0}\frac{\partial m^*}{\partial T}\neq0$, it
will conflict with the third law of thermodynamics,
$\lim_{T\rightarrow 0}S=0$. We hope to point out that this argument
is not right because it is based on Eq.(4) which has not taken the
medium effect into account. In Fig.4 we draw the entropy per baryon
vs. temperature curves for different baryon densities $n_B=0.2, 0.5,
1.0$ fm$^{-3}$ with dispersion relations
$m^*_1(T,n_B)=\frac{B_0(1-T/T_c)}{3n_B}$, using our treatment. We
even choose three different dispersion relations
$m^*_1(T,n_B)=\frac{B_0(1-T/T_c)}{3n_B}$,
$m^*_2(T,n_B)=\frac{B_0(1-T^2/T_c^{2})}{3n_B}$ and
$m^*_3(T,n_B)=\frac{B_0}{3n_B}$ for the same baryon density
$n_B=0.5fm^{-3}$ and find that all curves get together at the point
$T=0$, $S=0$. It means that our treatment is consistent with the
third law of thermodynamics no matter how the dispersion relation
is. The constraint Eq.(44) needs not to be satisfied.

\section{Summary}
In summary, we have shown the shortcomings of the previous
treatments, which based on the partial derivatives of
thermodynamical functions along a reversible process or based on the
additional terms to thermodynamical potential. The previous
treatments obstruct the correct consideration of medium in
thermodynamics. A new treatment of medium thermodynamics based on
equilibrium state is suggested. Employing QMDTD model, we address
the discrepancies between our treatment and others in Fig.1-4 and
Table I. We find that the negative pressure and the constraint
condition Eq.(44) are removed in our treatment.

\section*{Acknowledgements}
This work is supported in part by NNSF of
China and the National Basic Research Programme of China.

\begin{figure}[tbp]
\includegraphics[totalheight=16cm, width=16cm]{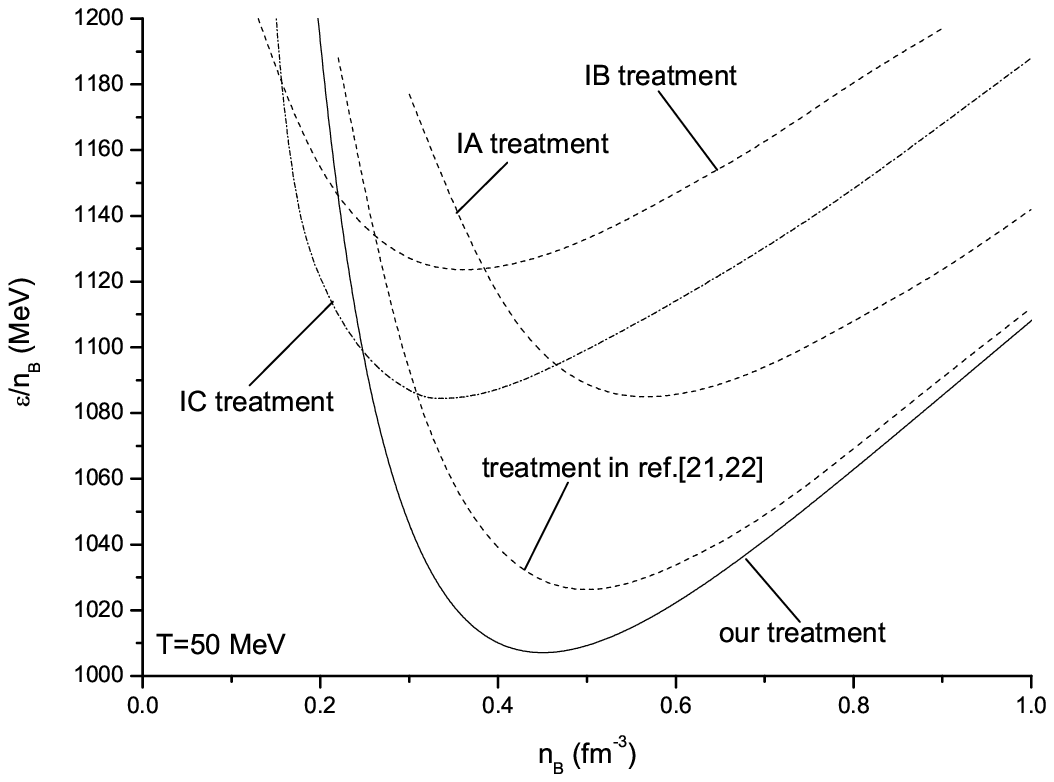}
\caption{Energy per baryon as a function of the baryon density
$n_{B}$ at $T=50$ MeV for different treatments \cite{Zhang:2002}.
The treatment IC and our treatment are for the QMDTD model, and
others are for the QMDD model.} \label{fig1}
\end{figure}

\begin{figure}[tbp]
\includegraphics[totalheight=16cm, width=16cm]{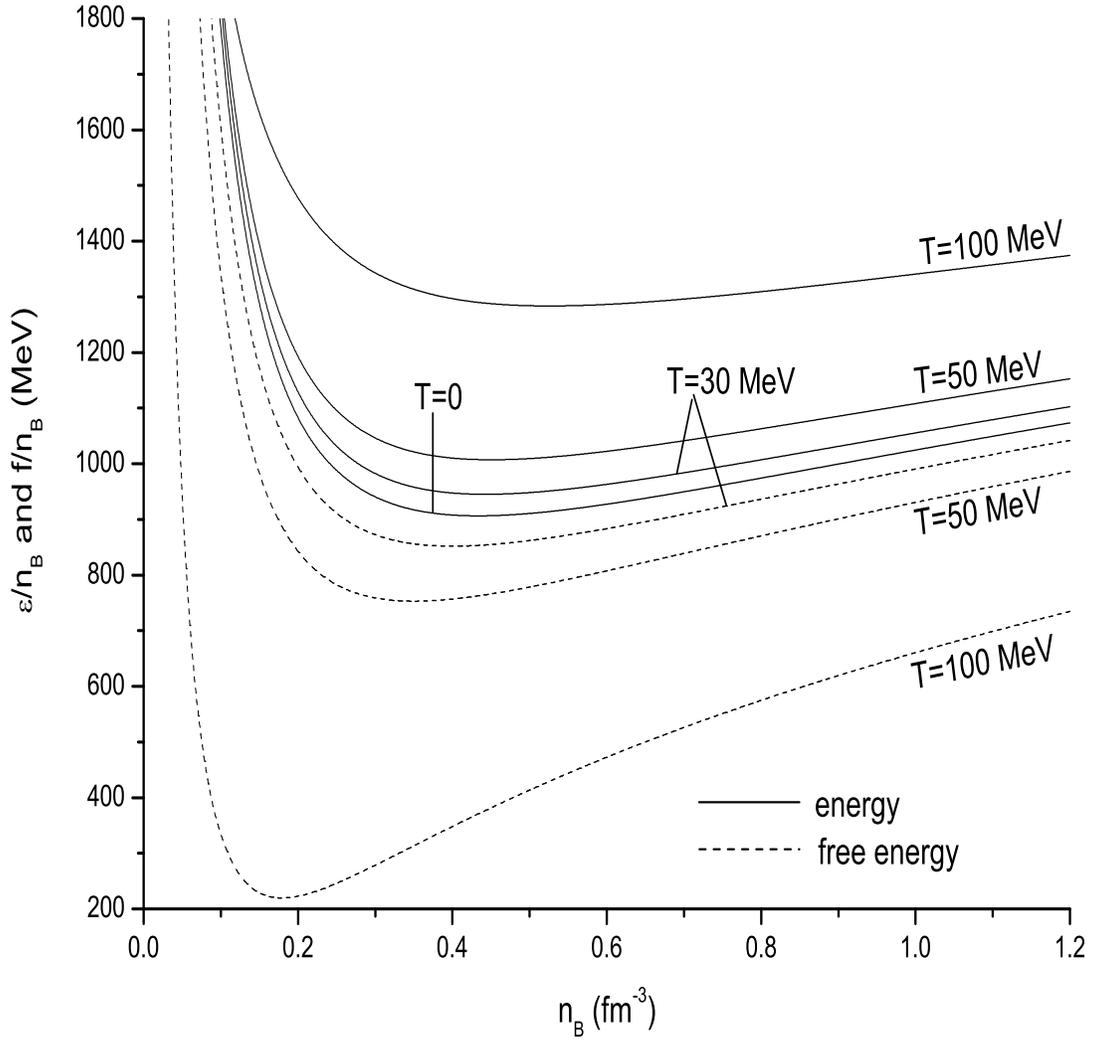}
\caption{Energy per baryon $U/N_B=\varepsilon/n_B$ and free energy
per baryon $F/N_B=f/n_B$ as functions of the baryon density $n_{B}$
at different temperatures $T=0, 30, 50, 100$ MeV. The two curves are
identical at $T=0$.} \label{fig2}
\end{figure}

\begin{figure}[tbp]
\includegraphics[totalheight=16cm, width=16cm]{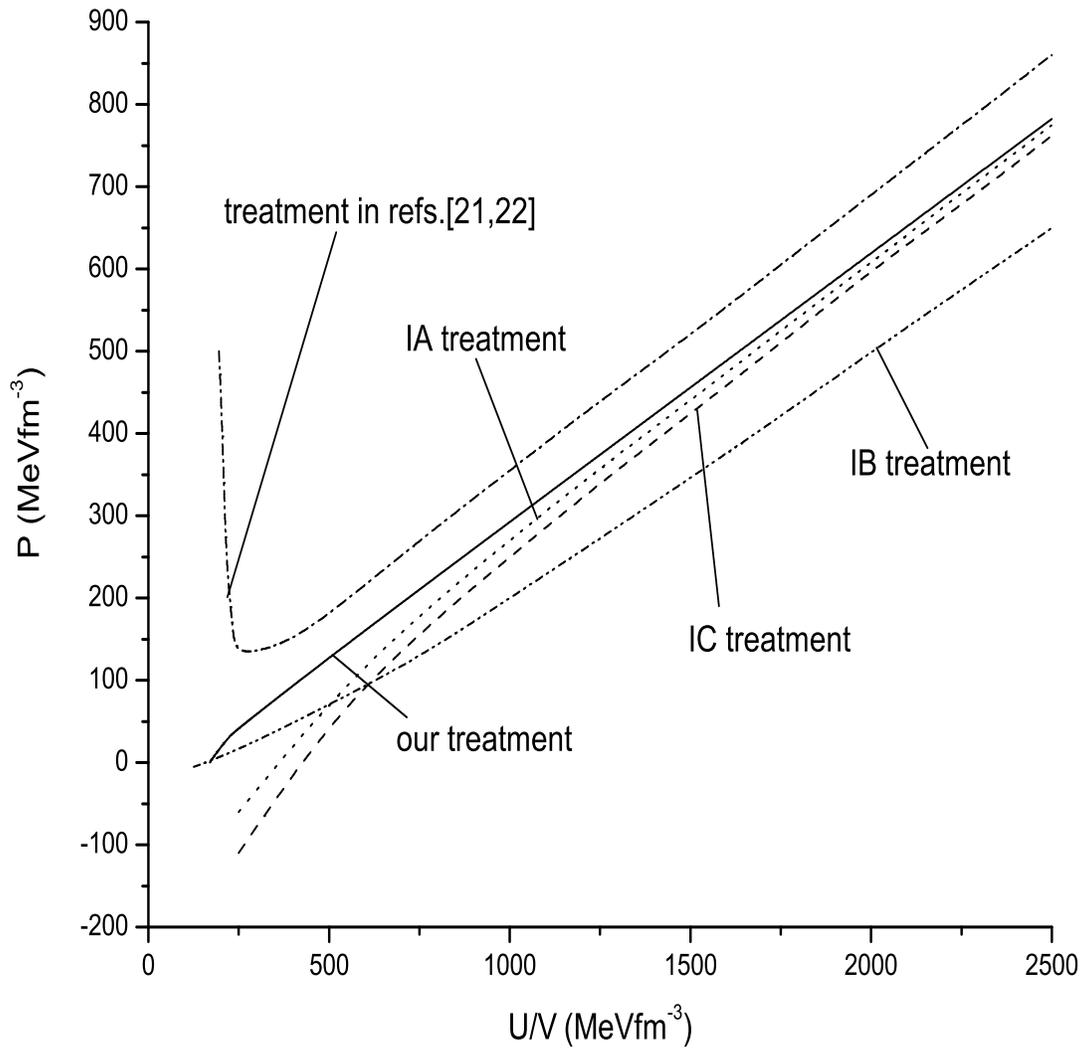}
\caption{Pressure $P$ as a function of the energy density $U/V$ for
different treatments of both the QMDD and the QMDTD models
\cite{Zhang:2002}. The tendencies of the curves are similar at large
energy density region, but at small energy density region, different
treatments have quite different behaviors. In our treatment, the
pressure never goes to negative.} \label{fig3}
\end{figure}

\begin{figure}[tbp]
\includegraphics[totalheight=16cm, width=16cm]{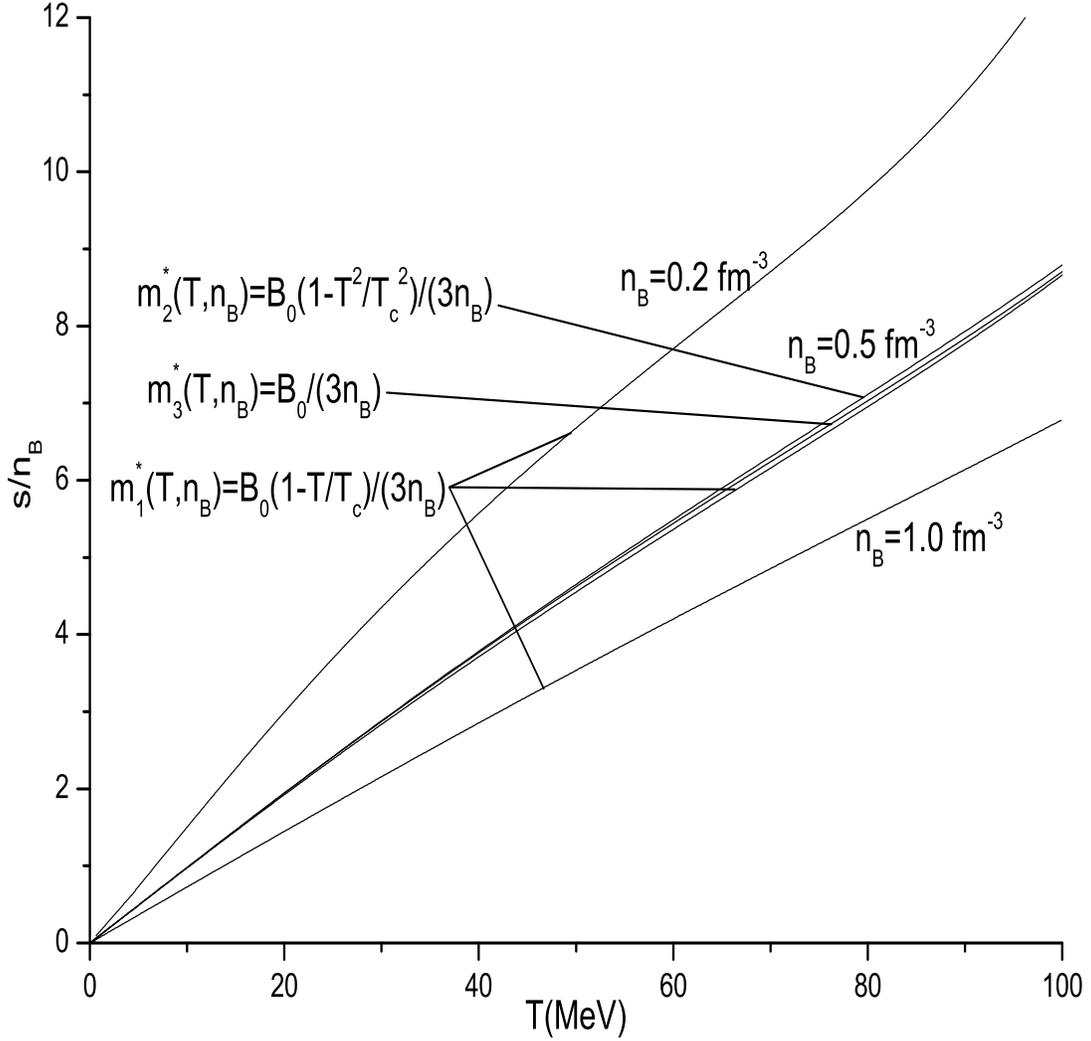}
\caption{Entropy per baryon vs. temperature curves for densities
$n_B=0.2, 0.5, 1.0$ fm$^{-3}$, where the mass has been taken as
$m_1^*(T,n_B)$, $m_2^*(T,n_B)$ and $m_3^*(T,n_B)$ respectively, as
indicated in the figure. All curves get together at $S=0$ when
$T=0$, which is consistent with the third law of thermodynamics.}
\label{fig5}
\end{figure}

\end{document}